\documentstyle{article}
\makeatletter

\def\case{\protect\@case}
\def\@case#1#2{%
\def\@tempa{#2}\def\@tempb{/}%
\ifx\@tempa\@tempb %
\def\@tempa{\@@case{#1}}%
\else %
\def\@tempa{\@@case{#1}{#2}}%
\fi
\@tempa}

\def\@@case#1#2{{\textstyle{#1\over#2}}}

\def\pmb#1{\leavevmode\setbox0=\hbox{#1}%
\kern-.02em\copy0\kern-\wd0
\kern.04em\copy0\kern-\wd0
\kern-.02em\raise.04em\box0 }

\def\lesssim{\mathrel{\mathpalette\vereq<}}
\def\vereq#1#2{\lower3pt\vbox{\baselineskip1.5pt \lineskip1.5pt
\ialign{$\m@th#1\hfill##\hfil$\crcr#2\crcr\sim\crcr}}}
\def\gtrsim{\mathrel{\mathpalette\vereq>}}

\def\tensor#1{\protect\@ontopof{#1}{\leftrightarrow}{1.15}\mathord{\box2}}
\def\overstar#1{\protect\@ontopof{#1}{\ast}{1.15}\mathord{\box2}}
\def\overdots#1{\protect\@ontopof{#1}{\cdots}{1.0}\mathord{\box2}}
\def\overcirc#1{\protect\@ontopof{#1}{\circ}{1.2}\mathord{\box2}}
\def\loarrow#1{\protect\@ontopof{#1}{\leftarrow}{1.15}\mathord{\box2}}
\def\roarrow#1{\protect\@ontopof{#1}{\rightarrow}{1.15}\mathord{\box2}}

\def\lambdabar{\protect\@lambdabar}
\def\@lambdabar{%
\relax
\bgroup
\def\@tempa{\hbox{\raise.73\ht0
\hbox to0pt{\kern.25\wd0\vrule width.5\wd0
height.1pt depth.1pt\hss}\box0}}%
\mathchoice{\setbox0\hbox{$\displaystyle\lambda$}\@tempa}%
{\setbox0\hbox{$\textstyle\lambda$}\@tempa}%
{\setbox0\hbox{$\scriptstyle\lambda$}\@tempa}%
{\setbox0\hbox{$\scriptscriptstyle\lambda$}\@tempa}%
\egroup}

\def\corresponds{{\lower.2ex\hbox{=}}{\rm\kern-.75em^\triangle}}
\def\succsim{\succ\kern-.9em_\sim\kern.3em}
\def\precsim{\prec\kern-1em_\sim\kern.3em}
\def\slantfrac#1#2{\kern1em^{#1}\kern-.3em/\kern-.1em_{#2}}

\def\mathhexbox{\protect\mathhexbox@}
\def\mathhexbox@#1#2#3{\relax
\ifmmode\mathpalette{}{\m@th\mathchar"#1#2#3}%
\else\leavevmode\hbox{$\m@th\mathchar"#1#2#3$}\fi}
\def\text#1{%
\relax
\ifmmode %
\mathchoice
{\hbox{\everymath{\displaystyle}\rm #1}}%
{\hbox{\everymath{\textstyle}\rm #1}}%
{\hbox{\everymath{\scriptstyle}%
\def\prm{\fam\z@ \the\scriptfont\z@ \relax}%
\def\pit{\fam\itfam \the\scriptfont\itfam \relax}%
\rm #1}%
}%
{\hbox{\everymath{\scriptscriptstyle}%
\def\prm{\fam\z@ \the\scriptscriptfont\z@ \relax}%
\def\pit{\fam\itfam \the\scriptscriptfont\itfam \relax}%
\rm #1}%
}%
\else %
\leavevmode\hbox{#1}%
\fi
}
\def\bbox#1{%
\leavevmode\text{%
\textfont0 \the\textfont\bffam
\scriptfont0 \the\scriptfont\bffam
\scriptscriptfont0 \the\scriptscriptfont\bffam
\@temptokena\everymath \boldmath \everymath\@temptokena
{$\m@th\relax#1$}%
}%
}

\def\[{\relax
\ifmmode\@badmath\else\par\vskip-\lastskip\vskip\abovedisplayskip\fi
\hbox to\hsize\bgroup
\def\label##1{\@revmess{warning}{\string\label{##1} used
in \string\[\space environment after (\theequation)}}%
\hskip\@totalleftmargin\hskip\@indentamount$\displaystyle}

\def\]{\relax
\ifmmode
$\hskip\@centering\egroup
\else
\@badmath
\fi
\vskip\belowdisplayskip
\global\@indentflag=1 %
\noindent\ignorespaces}

\newtoks\@stequation

\def\mathletters{\refstepcounter{equation}%
  \edef\@savedequation{\the\c@equation}%
  \@stequation=\expandafter{\theequation}
  \edef\@savedtheequation{\the\@stequation}
  \edef\oldtheequation{\theequation}%
  \setcounter{equation}{0}%
  \def\theequation{\oldtheequation\alph{equation}}}

\def\endmathletters{%
  \setcounter{equation}{\@savedequation}%
  \@stequation=\expandafter{\@savedtheequation}%
  \edef\theequation{\the\@stequation}%
  \global\@ignoretrue}
\newcount\@minsofar
\newcount\@min
\newcount\@cite@temp
\def\@citex[#1]#2{%
\if@filesw \immediate \write \@auxout {\string \citation {#2}}\fi
\@tempcntb\m@ne \let\@h@ld\relax \def\@citea{}%
\@min\m@ne%
\@cite{%
  \@for \@citeb:=#2\do {\@ifundefined {b@\@citeb}%
    {\@h@ld\@citea\@tempcntb\m@ne{\bf ?}%
    \@warning {Citation `\@citeb ' on page \thepage \space undefined}}%
{\@minsofar\z@ \@for \@scan@cites:=#2\do {%
  \@ifundefined{b@\@scan@cites}%
    {\@cite@temp\m@ne}
    {\@cite@temp\number\csname b@\@scan@cites \endcsname \relax}%
\ifnum\@cite@temp > \@min%
    \ifnum\@minsofar = \z@
      \@minsofar\number\@cite@temp
      \edef\@scan@copy{\@scan@cites}\else
    \ifnum\@cite@temp < \@minsofar
      \@minsofar\number\@cite@temp
      \edef\@scan@copy{\@scan@cites}\fi\fi\fi}\@tempcnta\@min
  \ifnum\@minsofar > \z@ 
    \advance\@tempcnta\@ne
    \@min\@minsofar
    \ifnum\@tempcnta=\@minsofar 
      \ifx\@h@ld\relax
        \edef \@h@ld{\@citea\csname b@\@scan@copy\endcsname}%
      \else \edef\@h@ld{\ifmmode{-}\else--\fi\csname b@\@scan@copy\endcsname}%
      \fi
    \else \@h@ld\@citea\csname b@\@scan@copy\endcsname
          \let\@h@ld\relax
  \fi 
\fi}%
\def\@citea{,\penalty\@highpenalty\,}}\@h@ld}{#1}}
\begin{document}
\title{{\bf Model independent analysis of the simultaneous mixing of gauge 
       bosons and mixing of fermions}\\
          \vspace{-14ex}
           \hfill{\normalsize hep-ph/9611442     \\}
           \hfill{\normalsize CINVESTAV-FIS-14/96\\}
          \vspace{11ex}}
\author{Umberto Cotti\protect\thanks{e-mail: ucotti@fis.cinvestav.mx}
       \, and \, Arnulfo Zepeda\\[2ex]
 \small \em  Departamento de F\'{\i}sica, Centro de Investigaci\'on y de
 \small \em  Estudios Avanzados del IPN, A.P. 14-740,\\
 \small \em  07000 M\'exico D.F., M\'exico.}
\date{\small (November, 1996)}
\maketitle

\vspace{-2ex}
\begin{abstract}
 We discuss the case of simultaneous mixing of gauge bosons and mixing 
of fermions in a model independent way and for a variety of 
extra-fermion representations. 
 In this context  we analyze a class of lepton family violating 
processes, namely 
$\rm Z \rightarrow e \bar{\tau}$,
$\rm Z \rightarrow \mu \bar{\tau}$, 
$\rm Z \rightarrow e \bar{\mu}$,
$\rm \mu \rightarrow ee\bar{e}$,
$\rm \tau \rightarrow ee\bar{e}$,
$\rm \tau \rightarrow \mu \mu \bar{\mu}$,
$\rm \tau \rightarrow e \mu \bar{\mu}$ and 
$\rm \tau \rightarrow \mu e \bar{e}$
in the presence of one extra neutral gauge boson, Z$'$, with universal, 
non-universal or family changing couplings. 
 We derive bounds on the combined effect of Z--Z$'$ mixing and
ordinary--exotic lepton mixing. 
\end{abstract}

\section{Introduction}
Tree level {\sl family changing neutral current} (FCNC) interactions arise 
in extended models from three possible sources: 
(i)~the exchange of family changing neutral gauge boson,
(ii)~the mixing between exotic and ordinary fermions and
(iii)~the existence of neutral scalars in the Higgs sector with family 
     violating couplings.
However, if the standard neutral $\rm Z^o$ boson mixes with a boson which 
has a coupling which is either family changing or nonuniversal, its coupling 
to the light (that is the ordinary) fermions becomes family changing 
even in the absence of mixing between exotic and ordinary fermions.

 In previous works an extensive research has been performed in the context
of FCNC produced by the mixing of the standard neutral gauge boson whith 
one which do not couple universally to fermion generation~\cite{prd32:306},
or by the mixing between exotic and ordinary 
fermions~\cite{prd38:886,npb386:239}.
 In this communication we show how this phenomenon arises in
the general case of simultaneous mixing of neutral 
gauge bosons and mixing of ordinary fermions with exotic ones. 
 We do not consider in this article FCNC arising from the exchange of
scalars, nor additional indirect effects such as the shifts induced by the 
mixing between the neutral gauge bosons in the values of the weak angle
$\rm \theta_w$, the $\rho$ parameter and the Fermi coupling constant
$\rm G_F$ \cite{prd46:3040,prd45:278,zpc53:97}, since they are irrelevant 
for the present analysis.
 We apply the formalism in a model independent way to several lepton 
family violating processes in the e--$\mu$, $\mu$--$\tau$ and 
e--$\tau$ sectors considering several possible exotic fermionic 
representations.
 We obtain in each case bounds for the mixing parameters  
including the possibility that the contribution of the neutral gauge
boson mixing and that of the fermion mixing are of the same order.

 We describe in section~\ref{mix-effgen} the formalism for dealing with FCNC 
which arise from simultaneous mixing of gauge bosons and mixing of fermions. 
 This formalism is applied to the leptonic sector in section~\ref{appl}.
 In subsection~\ref{Zlld} we describe how the mixing effects modify the 
diagonal couplings of the Z.
 In subsection~\ref{lfvp} we apply the formalism to the 
${\rm Z} \rightarrow l_i \bar{l}_j$,
$l_i \rightarrow l_j l_j \bar{l}_j$ and
$l_i \rightarrow l_j l_k \bar{l}_k$
decays and obtain constraints for the mixing parameters.
 These bounds are refined in section~\ref{irreps} considering special types of 
representations for the additional fermions.

\section{Mixing effects: the general formalism for simultaneous mixing 
         of gauge bosons and mixing of fermions}
\label{mix-effgen}
 To discuss the mixing of the massive  neutral gauge 
bosons of a general theory we first divide them  into two classes,
\begin{itemize}
 \item The standard Z$^{\rm o}$ gauge boson which is a linear combination of 
       the $\rm SU(2)_L \!\otimes\! U(1)_Y$~neutral bosons and has  
       {\sl universal family diagonal} (UFD) couplings 
       determined by the eigenvalues $t_{3}$ and $q$ of the 
       electroweak generators ${\sf T}_3$ and {\sf Q}.
 \item The extra Z$^{\rm o}_i$  gauge bosons which can have either UFD or 
       {\sl non universal family diagonal} NUFD or FC couplings. 
        The last two types of couplings arise when the Z$^{\rm o}_i$ gauge 
       bosons are associated with horizontal interactions.
 Since the case where Z$^{\rm o}_i$ has UFD couplings has already been 
discussed in the literature \cite{npb386:239,prd46:3040,plb248:139}, we 
will concentrate our attention on the cases of NUFD~\cite{prd32:306} and 
FC couplings.
\end{itemize}
 To discuss the general mixing of fermions, including additional ones,
we follow Langacker and London~\cite{prd38:886} grouping all fermions 
of a given electric charge, $q$, and a given helicity, 
$a$ = L, R, in a $n_a + m_a$ vector column of 
$n_a$ ordinary (O) and $m_a$ exotic (E) gauge eigenstates
$ \bbox{\psi}_a^{\rm o} = {\rm (\bbox{\psi}_O,\bbox{\psi}_E)}_a^{\top}$.
The relation between the gauge eigenstates
and the corresponding light ($l$) and heavy ($h$) mass eigenstates 
$\bbox{\psi}_a = (\bbox{\psi}_l,\bbox{\psi}_h)_a^{\top}$, is given by
\begin{equation}
 \bbox{\psi}_a^{\rm o} = {\sf U}_a \bbox{\psi}_a,
\label{fermass}
\end{equation}
where the unitary matrices {\sf U}$_a$ have the block form
\begin{equation}
 {\sf U}_a = 
  \left( 
   \begin{array}{cc}
    {\sf A}_a & {\sf E}_a \\[2mm]
    {\sf F}_a & {\sf G}_a  
   \end{array}  
  \right),
\end{equation}
and the submatrices ${\sf A}_a$ and ${\sf G}_a$ are not unitary but 
satisfy the following conditions
\begin{equation}
 \begin{array}{ccccc}
 \left(
  \sf U^\dagger \sf U
 \right)_a  & = & 
  \left( 
   \begin{array}{c@{\hspace{4mm}}c}
    {\sf A^{\dag}A + F^{\dag} F} & {\sf A^{\dag}E + F^{\dag}G} \\[2mm]
    {\sf E^{\dag}A + G^{\dag} F} & {\sf E^{\dag}E + G^{\dag}G}  
   \end{array}  
  \right)_a & = &
  \left( 
   \begin{array}{c@{\hspace{4mm}}c}
    {\sf 1} & {\sf 0} \\[2mm]
    {\sf 0} & {\sf 1}  
   \end{array}  
  \right) .
 \end{array}
\label{unit}
\end{equation}
The term $\left( \sf F^\dagger F \right)_a$, second order in the 
small exotic--ordinary
fermion mixing, induces FC transitions in the light--light sector. 

 The neutral current term for the multiplet $\bbox{\psi}$ of a given 
electric charge, for the case when both types of mixings are present, 
is then
\begin{mathletters}
\begin{eqnarray}
 -\cal{L}^{\rm nc} &=& 
  \frac{e}{\rm s_{\theta_w} c_{\theta_w}}
  \sum_{a={\rm L,R}} \bar{\bbox{\psi}}^{\rm o}_{a} 
  \gamma^{\mu} 
 \left(
  {\sf D}_a, {\sf H}^1_a, \cdots, {\sf H}^n_a  
 \right) 
 \bbox{\psi}^{\rm o}_{a}
  \left(
  \begin{array}{c}
   {\rm Z^o}         \\
   {\rm Z}^{\rm o}_1 \\
   \vdots            \\
   {\rm Z}^{\rm o}_{n} \\
  \end{array}
  \right)_{\mu} \label{general1} \\
  &=&
  \frac{e}{\rm s_{\theta_w} c_{\theta_w}}
  \sum_{a={\rm L,R}} \bar{\bbox{\psi}}_a \gamma^{\mu}
  \left(
   {\sf U}^{\dag}_a {\sf D}_a   {\sf U}_a, 
   {\sf U}^{\dag}_a {\sf H}^1_a {\sf U}_a, \cdots, 
   {\sf U}^{\dag}_a {\sf H}^n_a {\sf U}_a
  \right)_{\mu}
  \bbox{\psi}_a {\sf R}
  \left(
   \begin{array}{c}
     {\rm Z}  \\
    {\rm Z}_1 \\
    \vdots    \\
    {\rm Z}_n \\
   \end{array}
  \right)_{\mu} \label{general2}
\end{eqnarray}
where $\rm s_{\theta_w}$ and $\rm c_{\theta_w}$ are  
$\rm \sin \theta_w $ and $\rm \cos \theta_w$ respectively,
$\rm \theta_w$ is the weak mixing angle,
{\sf R} is the  $(n+1) \times (n+1)$ orthogonal matrix that diagonalizes 
the neutral boson
mass matrix, {\sf D}$_a$ is the $(n_a + m_a)\times(n_a + m_a)$  
matrix that expresses  the coupling of the 
$\rm Z^o$ gauge boson to matter fields, and similarly {\sf H}$^i_a$ are the 
$(n_a + m_a )\times(n_a + m_a)$  matrices that express
the coupling of the NUFD  and FC  
gauge bosons to matter.
\end{mathletters}
 The electromagnetic part of $\cal{L}^{\rm nc}$ has not been displayed
since its structure is not affected by the mixing effects.

 In the simple case of only one extra neutral gauge boson, the {\sf R} 
matrix is easily parametrizable as
\begin{eqnarray}
 {\sf R} 
 &=&
 \left(
  \begin{array}{cr}
    \cos{\Theta} & -\sin{\Theta}\\
   \sin{\Theta} & \cos{\Theta}
  \end{array}
 \right)
\end{eqnarray}
and the neutral current term is now,  
\begin{eqnarray}
 -\cal{L}^{\rm nc} &=& 
  \frac{e}{\rm s_{\theta_w} c_{\theta_w}}
 \sum_{a={\rm L,R}} \bar{\bbox{\psi}}_a \gamma^{\mu} 
 {\sf U}^{\dag}_a 
 \left( 
  {\sf D}_a   \cos{\Theta} + 
  {\sf H}_a \sin{\Theta} , 
  {\sf H}_a \cos{\Theta} - 
  {\sf D}_a    \sin{\Theta}  
 \right) 
 {\sf U}_a
 \bbox{\psi}_a
 \left(
  \begin{array}{l}
   {\rm Z}  \\
   {\rm Z}' \\
  \end{array}
 \right)_{\mu} . \label{ncterm}
\end{eqnarray} 
 It should be obvious that the treatment of the general case, 
eqs.~(\ref{general1}) and~(\ref{general2}), is straightforward.
 From now on we restrict the discussion to the case of only one 
extra gauge boson just to simplify the notation.
 In eqs.~(\ref{general1}), (\ref{general2}) and~(\ref{ncterm}) 
the ${\sf D}_a$ matrices are given by
\begin{eqnarray}
 {\sf D}_a  &\equiv&
  \left(
   {\sf T}_{3} - {\sf Q}{\rm s^2_{\theta_w}}
  \right)_a.
\end{eqnarray}
They are diagonal by definition and in the case, which we assume for 
simplicity from now on, that the exotics of a given charge and helicity 
have a common eigenvalue $t_{{\rm 3E}a}$ of ${\sf T}_{3a}$, they can be 
written as 
\begin{eqnarray}
{\sf D}_a  &=&
  \left(
   \begin{array}{cc}
    {\sf t}_{\rm 3O} 
     - {\sf q}_{\rm O}{\rm s^2_{\theta_w}} & 
{\sf 0}\\
    {\sf 0} & {\sf t}_{\rm 3E} 
     - {\sf q}_{\rm E}{\rm s^2_{\theta_w}}
   \end{array}
  \right)_a ,
\end{eqnarray}
where ${\sf t}_{{\rm 3O}a}$ and ${\sf t}_{{\rm 3E}a}$ are square matrices
of dimension $n_a$ and $m_a$ respectively.
 They correspond to the ordinary and exotic part of the ${\sf T}_{3a}$
operator and they are proportional to the unit matrix through the 
eigenvalues $t_{{\rm 3O}a}$ and $t_{{\rm 3E}a}$ of ${\sf T}_{3a}$.
 This is the same situation for the ${\sf q}_{{\rm O}a}$ and
${\sf q}_{{\rm E}a}$ matrices in relation with the {\sf Q} operator.

 Contrary to the {\sf D}$_a$ matrices, the {\sf H} ones are not diagonal 
in general but can however be written as 
\begin{equation}
 {\sf H} = 
  \left(
   \begin{array}{cc}
    {\sf H}_{\rm O} &    {\sf 0} \\
       {\sf 0}      & {\sf H}_{\rm E} 
   \end{array}
  \right),
\end{equation} 
where {\sf H}$_{\rm O}$ and {\sf H}$_{\rm E}$ represent the interactions of
the $\rm Z^o_1$ with the ordinary and exotics fermions respectively.
 The point is that there are no {\sf H}$_{\rm EO}$ nor 
{\sf H}$_{\rm OE}$ terms in ${\sf H}$ (which would give rise to $\rm Z^o_1$
mediated transitions between exotic and ordinary fermions) as long as 
the horizontal group commutes with the Standard Model (SM) gauge group.

 In the basis where the fermions are mass eigenstates, the form of {\sf D} 
and {\sf H} is 
\begin{mathletters}
\begin{eqnarray}
  \left( 
   {\sf U^{\dag} D U}
  \right)_{\rm R} &=&
  \left(
   \begin{array}{cc}
    {\sf F^{\dag} F} & {\sf F^{\dag} G} \label{udur} \\[2mm]
    {\sf G^{\dag} F} & {\sf G^{\dag} G} 
   \end{array}
  \right)_{\rm R} 
   t_{\rm 3ER} - {\sf Q}{\rm s^2_{\theta_w}} ,  \label{DR}\\[3mm]
 \left( 
  {\sf U^{\dag} D U}
 \right)_{\rm L} &=&
  \left(
    \begin{array}{cc}
     {\sf F^{\dag} F} & {\sf F^{\dag} G} \\[2mm]
     {\sf G^{\dag} F} & {\sf G^{\dag} G} 
    \end{array}
  \right)_{\rm L} 
   t_{\rm 3EL} +
  \left(
   \begin{array}{cc}
    {\sf A^{\dag} A} & {\sf A^{\dag} E} \\[2mm]
    {\sf E^{\dag} A} & {\sf E^{\dag} E} 
   \end{array}
  \right)_{\rm L} 
   t_{\rm 3OL} - {\sf Q}{\rm s^2_{\theta_w}},
\end{eqnarray}
and by using the unitarity conditions~(\ref{unit}) we can rewrite the last 
equation as
\begin{eqnarray}
 \left( 
  {\sf U^{\dag} D U}
 \right)_{\rm L} &=&
  \left(
   \begin{array}{cc}
    {\sf F^{\dag} F} & -{\sf A^{\dag} E} \\[2mm]
    {\sf G^{\dag} F} & -{\sf E^{\dag} E} 
   \end{array}
  \right)_{\rm L} 
   (t_{\rm 3EL} - t_{\rm 3OL}) + {\sf T}_{\rm 3L} - 
   {\sf Q}{\rm s^2_{\theta_w}}. 
\label{DL}
\end{eqnarray}
 From eq.~(\ref{udur}) one can see that for the light fermions, and in 
the absence of $\rm Z^o$--$\rm Z^o_1$ mixing, the coupling of the 
$\rm Z^o$ to right handed FC and 
{\sl non-universal family diagonal neutral currents} (NUFDNC) is possible 
only if $t_{\rm 3ER} \neq {\sf 0}$. 
 Furthermore, from eqs.~(\ref{DR}) and (\ref{DL}) it's easy to see that 
sequential fermions 
do not induce FC nor NUFD couplings for the standard $\rm Z^o$ since 
their contribution to these currents is canceled out by that of the 
ordinary fermions.
\end{mathletters}

 On the other hand, no general statement can be made for the 
transformed {\sf H} couplings:
\begin{eqnarray}
 \left(
  {\sf U^{\dag} H U}
 \right)_a
  &=&
  \left(
   \begin{array}{cc}
    {\sf A^{\dag}} {\sf H}_{\rm O} {\sf A} \;+\; 
      {\sf F^{\dag}} {\sf H}_{\rm E} {\sf F}, &
    {\sf A^{\dag}} {\sf H}_{\rm O} {\sf E} \;+\;
      {\sf F^{\dag}} {\sf H}_{\rm E} {\sf G} \\[2mm]
    {\sf E^{\dag}} {\sf H}_{\rm O} {\sf A} \;+\;
      {\sf G^{\dag}} {\sf H}_{\rm E} {\sf F}, &
    {\sf E^{\dag}} {\sf H}_{\rm O} {\sf E} \;+\;
      {\sf G^{\dag}} {\sf H}_{\rm E} {\sf G}
   \end{array}
  \right)_a 
  \equiv \;
  \left(
   \begin{array}{cc}
    {\sf H}_{ll}  & {\sf H}_{lh} \\[2mm]
    {\sf H}^{\dag}_{lh} & {\sf H}_{hh}
   \end{array}
  \right)_a . \label{H}
\end{eqnarray}
 From the last equation one can see that in the presence of neutral 
gauge boson mixing, $\Theta \neq 0$, there will be in general FC 
couplings of the $\rm Z$ in the light sector, ${\sf H}_{ll}$ nondiagonal, 
even in the absence of mixing between exotic and ordinary fermions, 
${\sf F} = {\sf 0}$ and even if the coupling of $\rm Z^o_1$ to the ordinary 
fermions, ${\sf H}_{\rm O}$, is diagonal but nonuniversal, since in general 
the mass and gauge eigenstates will not coincide in the light sector, 
${\sf A}\neq 1$.

\subsection{The general neutral current lagrangian term in the light 
            sector}
 From eqs. (\ref{DR}), (\ref{DL}) and (\ref{H}) we 
obtain for the neutral-current lagrangian in the light--light 
sector the following expression:
\begin{eqnarray}
 -\cal{L}^{\rm nc} & = &
 \frac{e}{\rm s_{\theta_w} c_{\theta_w}} \sum_{a={\rm L,R}} 
    \bar{\bbox{\psi}}_{la} \gamma^{\mu}
 \left(
  {\sf K}_a,\; {\sf K'}_a
 \right)
 \bbox{\psi}_{la}
 \left(
  \begin{array}{l}
   {\rm Z}  \\
   {\rm Z}' \\
  \end{array}
 \right)_{\mu}.
\label{lnc}
\end{eqnarray}
where
\begin{mathletters}
\label{coeff}
\begin{eqnarray}
 {\sf K}_{\rm L} & = &
 \left[
  \left( {\sf F^{\dag} F} \right)_{\rm L}(t_{\rm 3EL} - t_{\rm 3OL})
   + {\sf t}_{\rm 3OL} - {\sf Q}{\rm s_{\theta_w}^2} 
 \right]
 \cos{\Theta} +
 \left( 
  {\sf H}_{ll} 
 \right)_{\rm L} \sin{\Theta}, \label{kl1}\\[3mm]
 {\sf K}_{\rm R} & = &  
 \left[
  \left(
   {\sf F^{\dag} F} \right)_{\rm R}t_{\rm 3ER} - 
   {\sf Q}{\rm s_{\theta_w}^2}
 \right]
 \cos{\Theta} +
 \left( 
  {\sf H}_{ll} 
 \right)_{\rm R}\sin{\Theta}, \label{kr1}\\[3mm]
 {\sf K}^{\prime}_{\rm L} & = & -
 \left[
  \left(
   {\sf F^{\dag} F} 
  \right)_{\rm L}
  \left( 
   t_{\rm 3EL} - t_{\rm 3OL}
  \right) 
   + {\sf t}_{\rm 3OL} - {\sf Q}{\rm s_{\theta_w}^2} 
 \right]
 \sin{\Theta} +
 \left(
  {\sf H}_{ll} 
 \right)_{\rm L} \cos{\Theta}, \\[3mm]
 {\sf K}^{\prime}_{\rm R} & = &-
 \left[
  \left( 
   {\sf F^{\dag} F}
  \right)_{\rm R}t_{\rm 3ER} - {\sf Q}{\rm s_{\theta_w}^2}
 \right]
 \sin{\Theta} +
 \left( 
  {\sf H}_{ll} 
 \right)_{\rm R}\cos{\Theta}.
\end{eqnarray}
 From these eqs. it is easy to see that:
\end{mathletters}
\begin{sloppypar}
\begin{itemize}
\item[--] There are two contributions to the FC couplings of the 
 light fermions to the  Z, proportional to 
 $\left( {\sf F^{\dag} F} \right)_a \cos{\Theta}$ and 
 $\left( {\sf H}_{ll}\right)_a \sin{\Theta}$ respectively, 
 which may be in principle of the same order;
\item[--] In the limit of no mixing between exotics and ordinary fermions 
 $\left( {\sf F}_a = {\sf 0}\right)$ and no mixing between the Z and 
 the extra gauge boson $\left( \Theta = 0\right)$ the SM couplings are 
 recovered;
\item[--] In the absence of mixing with the exotic fermions, the FC 
 couplings of the ordinary fermions (of a given helicity) to the Z may still 
 survive through the term $\left( {\sf H}_{ll}\right)_a \sin{\Theta}$, 
 provided that the family of ordinary fermions of the given helicity
 transforms nontrivially under the horizontal generator ${\sf H}_{\rm O}$.
\end{itemize}
\end{sloppypar}
Further details of these couplings depend on the model and on the processes
under consideration and are the subject of the next sections.

 We may rewrite eqs.~(\ref{kl1}) and (\ref{kr1}) as
\begin{mathletters}
\label{coeff2}
\begin{eqnarray}
 {\sf K}_{\rm L} & = &
 \left(
  \bbox{\Lambda}_{\rm L} + {\sf t}_{\rm 3OL} - {\sf Q}{\rm s_{\theta_w}^2}
 \right) \cos{\Theta} +
 \bbox{\Xi}_{\rm L}\sin{\Theta}, \label{kl}\\[3mm]
 {\sf K}_{\rm R} & = & 
 \left( 
  \bbox{\Lambda}_{\rm R} - {\sf Q}{\rm s_{\theta_w}^2} 
 \right) \cos{\Theta} +
 \bbox{\Xi}_{\rm R}\sin{\Theta}, \label{kr}
\end{eqnarray}
where
\end{mathletters}
\begin{mathletters}
\begin{eqnarray}
 \bbox{\Lambda}_{\rm L} &=& 
  \left( 
   {\sf F^{\dag} F} 
  \right)_{\rm L}(t_{\rm 3EL} - t_{\rm 3OL}), \\[3mm]
 \bbox{\Lambda}_{\rm R} &=& 
  \left( 
   {\sf F^{\dag} F} 
  \right)_{\rm R} t_{\rm 3ER}, \\[3mm]
 \bbox{\Xi}_a &=& 
 \left( 
  {\sf H}_{ll} 
 \right)_a,
\end{eqnarray}
together with $\Theta$, represent the physics beyond the SM.
\end{mathletters}

\subsubsection{Charged fermions}
Since for the light charged fermions, the dimension of 
$\bbox{\psi}_{l{\rm L}}$ and $\bbox{\psi}_{l {\rm R}}$ are the same 
(there is an equal number of left and right handed fermions), we can rewrite 
the general lagrangian~(\ref{lnc}) as  
\begin{eqnarray}
 -\cal{L}^{\rm nc} & = &
  \frac{e}{2\rm s_{\theta_w} c_{\theta_w}} 
    \bar{\bbox{\psi}}_{l} \gamma^{\mu}
 \left(
    {\sf g}_{\rm V} - {\sf g}_{\rm A} \gamma^5,\; 
    {\sf g}^{\prime}_{\rm V} - {\sf g}^{\prime}_{\rm A} \gamma^5
 \right)
 \bbox{\psi}_{l}
 \left(
  \begin{array}{l}
   {\rm Z}  \\
   {\rm Z}' \\
  \end{array}
 \right)_{\mu},
\end{eqnarray}
where
\begin{mathletters}
\begin{eqnarray}
 {\sf g}_{\rm V} &=& {\sf K _{\rm L} + K_{\rm R}}, \\
 {\sf g}_{\rm A} &=& {\sf K _{\rm L} - K_{\rm R}}, 
\end{eqnarray}
\end{mathletters}

\section{Applications to the leptonic sector}
\label{appl}

\subsection{Constraints from the lepton family diagonal processes
            $\bbox{{\rm Z} \rightarrow l_i \bar{l}_i}$}
\label{Zlld}
 The effects of mixing between ordinary and and exotic fermions on the 
diagonal process ${\rm Z} \rightarrow l_i \bar{l}_i$ has been analyzed 
previously~\cite{prd38:886,npb386:239,prd46:3040}.
 Likewise separate effect of mixing between the standard Z and a new one
were discussed in Ref.~\cite{prd46:3040,prd45:278}.
 When both effects are present, the branching ratio 
${\rm B}({\rm Z} \rightarrow l_i \bar{l}_i)$, in the 
$M_{\rm Z} \gg m_{l_i}$ approximation, is given by
\begin{mathletters}
\begin{eqnarray}
 {\rm B}({\rm Z} \rightarrow l_i \bar{l}_i) &\simeq&
 \frac{1}{\Gamma_{\rm tot}} \frac{{\rm G_F} M^3_{\rm Z}}{6 \sqrt{2} \pi}
 \left(
  \left| 
   g_{\rm V}^{ii} 
  \right|^2 +  
  \left| 
   g_{\rm A}^{ii} 
  \right|^2
 \right) \\[3mm]
 &=&
 \frac{1}{\Gamma_{\rm tot}} \frac{{\rm G_F} M^3_{\rm Z}}{3 \sqrt{2} \pi}
 \left(
  \left| 
   \Lambda_{\rm L}^{ii} + \Xi_{\rm L}^{ii} \Theta - 
     \case{1}{2} + {\rm s_{\theta_w}^2}
  \right|^2 +  
  \left|  
   \Lambda_{\rm R}^{ii} + \Xi_{\rm R}^{ii} \Theta + {\rm s_{\theta_w}^2}   
  \right|^2
 \right)  + {\rm O}(\Theta^2).
\end{eqnarray}
 Since the agreement of the SM predictions with the experimental data 
for these processes is better than 0.1~\% 
(the experimental value of $\Gamma({\rm Z} \rightarrow l\bar{l})$ is 
$83.83 \pm 0.27$~\cite{prd54:1} against the theoretical one equal to 
$83.97 \pm 0.07$), the quantities $\Lambda_a^{ii} + \Xi_a^{ii} \Theta$ 
are bounded practically by the experimental uncertainty in the 
data~\cite{prd54:1},
\end{mathletters}
\begin{eqnarray}
 {\rm B}_{l_i \bar{l}_i} \equiv
  {\rm B}({\rm Z} \rightarrow l_i \bar{l}_i) &=& 
 \left\{
  \begin{array}{lcl}
   \rm B_{e \bar{e}}       &=& (3.366 \pm 0.008)  \times  10^{-2} \\[2mm]
   \rm B_{\mu \bar{\mu}}   &=& (3.367 \pm 0.013)  \times  10^{-2} \\[2mm]
   \rm B_{\tau \bar{\tau}} &=& (3.360 \pm 0.015)  \times  10^{-2} .
  \end{array}
 \right.
\end{eqnarray}
We may also write
\begin{eqnarray}
\label{values1}
 \left| 
  \Lambda_{\rm L}^{ii} + \Xi_{\rm L}^{ii} \Theta -
    \case{1}{2} + {\rm s_{\theta_w}^2}
 \right|^2 +
 \left| 
  \Lambda_{\rm R}^{ii} + \Xi_{\rm R}^{ii} \Theta + {\rm s_{\theta_w}^2}
 \right|^2
 &=& c {\rm B}_{l_i \bar{l}_i},
\end{eqnarray}
where 
$c^{-1} = \left( \frac{1}{\Gamma_{\rm tot}} 
\frac{{\rm G_F} M^3_{\rm Z}}{3 \sqrt{2} \pi}\right) = 0.2675 \pm 0.0005$ 
and from which we obtain, in a neighborhood of 
$\left| \Lambda_a^{ii} + \Xi_a^{ii} \Theta \right| = 0$ 
and with $\rm s^2_{\theta_w} = 0.2237 \pm 0.0010$,
the bounds
\begin{eqnarray}
 \left| 
  \Lambda_a^{ii} + \Xi_a^{ii} \Theta 
 \right| &<& {\rm few} \times 10^{-3} .
 \label{few}
\end{eqnarray}.

\subsection{Constraints from lepton family violating processes}
\label{lfvp}

\subsubsection{Constraints from ${\rm Z} \rightarrow l_i \bar{l}_j$}
\label{Zll}
 With the approximation $M_{\rm Z} \gg m_{l_i}, m_{l_j}$, and 
taking into account that experimental limits exists only for the sum of 
the charge states of particles and antiparticles states, we should consider 
for $i \neq j$ 
\begin{mathletters}
\begin{eqnarray}
 {\rm B}({\rm Z} \rightarrow l_i \bar{l}_j + \bar{l}_i l_j) &\simeq&
 2\frac{{\rm B}({\rm Z} \rightarrow l \bar{l})}
 {\left|
   g_{\rm V}
  \right|^2 + 
  \left|
   g_{\rm A}
  \right|^2}
 \left( 
  \left| 
   g_{\rm V}^{ij} 
  \right|^2 + 
  \left| 
   g_{\rm A}^{ij} 
  \right|^2 
 \right) \\[3mm]
&\simeq&
  4\frac{{\rm B}({\rm Z} \rightarrow 
                  l \bar{l})}{\left|
                               g_{\rm V}
                              \right|^2 + 
                              \left|
                               g_{\rm A}
                              \right|^2} 
 \left(
  \left|
   \Lambda_{\rm L}^{ij} + \Xi_{\rm L}^{ij} \Theta
  \right|^2 +
  \left|
   \Lambda_{\rm R}^{ij} + \Xi_{\rm R}^{ij} \Theta
  \right|^2
 \right) + {\rm O}\left( \Theta^2 \right).
\end{eqnarray}
 It then follows that 
\end{mathletters}
\begin{eqnarray}
\label{bounds1}
 \left| \Lambda_{\rm L}^{ij} + \Xi_{\rm L}^{ij} \Theta \right|^2 +
 \left| \Lambda_{\rm R}^{ij} + \Xi_{\rm R}^{ij} \Theta \right|^2
 &<& c {\rm \widetilde{\rm B}}_{l_i \bar{l}_j},
\end{eqnarray}
where $c^{-1} = \left(4\frac{{\rm B}({\rm Z} \rightarrow l \bar{l})}
  {\left| g_{\rm V}\right|^2 + \left|g_{\rm A}\right|^2} \right) = 0.536$
(using the conventional SM branching ratio 0.0337 for B$_{l\bar{l}}$
and the standard values for $g_{\rm V}$ and $g_{\rm A}$) 
and where 
\begin{eqnarray}
 {\rm B}_{l_i \bar{l}_j} \equiv
   {\rm B}({\rm Z} \rightarrow l_i \bar{l}_j + \bar{l}_i l_j) 
    &=& 
 \left\{
  \begin{array}{ccccl}
   \rm B_{e \bar{\mu}}    &<& 1.7 \times 10^{-6} &\equiv& 
    \rm \widetilde{B}_{e \bar{\mu}}                         \\[2mm]
   \rm B_{e \bar{\tau}}   &<& 7.3 \times 10^{-6} &\equiv& 
    \rm \widetilde{B}_{e \bar{\tau}}                        \\[2mm]
   \rm B_{\mu \bar{\tau}} &<& 1.0 \times 10^{-5} &\equiv&
    \rm \widetilde{B}_{\mu \bar{\tau}},
  \end{array}
 \right.
\end{eqnarray}
according to the experimental
limits~\cite{zpC67:555,L3:1798}.
 This means that the fermion mixing parameters 
$\Lambda_a^{ij}$ are bounded
to lie in a circular region centered at
($-\Xi_{\rm L}^{ij} \Theta$, $-\Xi_{\rm R}^{ij} \Theta$ ) and of 
radius $\sim 10^{-3}$.

 It's evident that the contribution of $\Theta$ in the analysis of
$\Lambda_a^{ij}$ is non negligible when 
\begin{equation}
 \Xi_a^{ij}\Theta \gtrsim \sqrt{{\rm c}\widetilde{\rm B}_{l_i \bar{l}_j}}.
\end{equation}
This may be a common situation, since in general
$\Xi_a^{ij}$ = O(1) and the upper bounds 
for $\Theta$, which are model dependent, are of the order of
$10^{-1}$ to $10^{-3}$.
 Taking the limit $\Theta \rightarrow 0$ could lead to wrong conclusions:
a contribution of $\Xi_a^{ij} \Theta \sim 5\times 10^{-3}$ is enough 
to give a completely new region of solutions for $\Lambda_a^{ij}$.
 The results of this section are resumed in table \ref{table1}.

\subsubsection{Constraints from $l_i \rightarrow l_j l_j \bar{l}_j$}
\label{liljlj}
Assuming $m_{l_i} \gg m_{l_j}$ and ignoring possible contributions from 
scalars,
the branching ratio ${\rm B}(l_i \rightarrow l_j l_j \bar{l}_j)$ for 
$i \neq j$ is
\begin{mathletters}
\begin{eqnarray}
 \frac{{\rm B} (l_i \rightarrow l_j l_j \bar{l}_j)}
      {{\rm B} (l_i \rightarrow l_j \bar{\nu}_{l_j} \nu_{l_i})} &=&
  \frac{1}{2}
  \left[ 3
   \left(
    \left|g^{jj}_{\rm V}\right|^2 + \left|g^{jj}_{\rm A}\right|^2
   \right) 
   \left(
    \left|g^{ij}_{\rm V}\right|^2 + \left|g^{ij}_{\rm A}\right|^2
   \right) + \right. \nonumber \\[2mm] && \left. \hspace{18mm}
   2\Re e
   \left(
    g^{jj}_{\rm V} {g^{jj}_{\rm A}}^\ast
   \right)   
   2\Re e
   \left(
    g^{ij}_{\rm V} {g^{ij}_{\rm A}}^\ast
   \right)
  \right] + \nonumber \\ &&
   \frac{M^2_{\rm Z}}{M^2_{\rm Z'}}
  \Re e
  \left[
  3
  \left(
   g^{jj}_{\rm V} {g^{\prime jj}_{\rm V}}^\ast + 
   g^{jj}_{\rm A} {g^{\prime jj}_{\rm A}}^\ast
  \right)
  \left(
   g^{ij}_{\rm V} {g^{\prime ij}_{\rm V}}^\ast + 
   g^{ij}_{\rm A} {g^{\prime ij}_{\rm A}}^\ast
  \right) + \right. \nonumber \\[2mm] && \left. \hspace{18mm}
  \left(
   g^{jj}_{\rm V} {g^{\prime jj}_{\rm A}}^\ast + 
   g^{jj}_{\rm A} {g^{\prime jj}_{\rm V}}^\ast
  \right)
  \left(
   g^{jj}_{\rm V} {g^{\prime jj}_{\rm A}}^\ast + 
   g^{jj}_{\rm A} {g^{\prime jj}_{\rm V}}^\ast
  \right)
  \right] + \nonumber \\ &&
   \frac{1}{2} \frac{M^4_{\rm Z}}{M^4_{\rm Z'}}
  \left[ 3
   \left(
    \left|g^{\prime jj}_{\rm V}\right|^2 + \left|g^{\prime jj}_{\rm A}\right|^2
   \right) 
   \left(
    \left|g^{\prime ij}_{\rm V}\right|^2 + \left|g^{\prime ij}_{\rm A}\right|^2
   \right) + \right. \nonumber \\[2mm] && \left. \hspace{18mm}
   2\Re e
   \left(
    g^{\prime jj}_{\rm V} {g^{\prime jj}_{\rm A}}^\ast
   \right)   
   2\Re e
   \left(
    {g^{\prime ij}_{\rm V}} {g^{\prime ij}_{\rm A}}^\ast
   \right)
  \right] 
\label{explain} \\[3mm]
&\simeq&
 4\left[
  \left(
   2\left|
    -\case{1}{2} + {\rm s_{\theta_w}^2}
   \right|^2 +
   \left|
    {\rm s_{\theta_w}^2}
   \right|^2 
  \right)
   \left|
    \Lambda^{ij}_{\rm L} + \Xi^{ij}_{\rm L} \Theta 
   \right|^2 + \right. \nonumber \\[2mm] && \left. \hspace{18mm}
  \left(
   \left|
    -\case{1}{2} + {\rm s_{\theta_w}^2}
   \right|^2 +
   2\left|
    {\rm s_{\theta_w}^2}
   \right|^2 
  \right)
   \left|
    \Lambda^{ij}_{\rm R} + \Xi^{ij}_{\rm R} \Theta 
   \right|^2   
 \right]  + {\rm O}\left( \Theta^2\right),
\end{eqnarray} 
 where we have assumed
$\left(\frac{M_{\rm Z}}{M_{\rm Z'}}\right)^2 \sim \Theta$,
$\Lambda_a^{ij} \lesssim \Theta$
(remember that $\Lambda_a^{ij}$ is second order in the ordinary--exotic mixing)
and we have taken into account the stringent
limits obtained in eq.~(\ref{few}) from which 
\end{mathletters}
\begin{mathletters}
\begin{eqnarray}
 \left|
  \Lambda_{\rm L}^{jj} + \Xi^{jj}_{\rm L} \Theta
    -\case{1}{2} + {\rm s_{\theta_w}^2}
 \right| &\simeq&
 \left|
  -\case{1}{2} + {\rm s_{\theta_w}^2}
 \right|, \\[3mm]
 \left|
  \Lambda_{\rm R}^{jj} + \Xi^{jj}_{\rm R} \Theta + {\rm s_{\theta_w}^2}
 \right| &\simeq&
 \left|
  {\rm s_{\theta_w}^2}
 \right|.
\end{eqnarray}
 Using the experimental bounds~\cite{npb299:1,prl73:1890,zpc55:179}
\end{mathletters}
\begin{equation}
 {\rm B}_{l_i l_j l_j \bar{l}_j} \equiv
 {\rm B}(l_i \rightarrow l_j l_j \bar{l}_j ) = 
 \left\{
  \begin{array}{lcclcc}
   \rm B_{\mu e e \bar{e}}        &<& 1.0 \times 10^{-12} &\equiv& 
                              \rm \widetilde{B}_{\mu e e \bar{e}}      \\[2mm]
   \rm B_{\tau e e \bar{e}}       &<& 3.3 \times 10^{-6}  &\equiv& 
                              \rm \widetilde{B}_{\tau e e \bar{e}}     \\[2mm]
   \rm B_{\tau \mu \mu \bar{\mu}} &<& 1.9 \times 10^{-6}  &\equiv&
                              \rm \widetilde{B}_{\tau \mu \mu \bar{\mu}}
  \end{array}
 \right.
\end{equation}
and 
$\rm s^2_{\theta_w} = 0.2237$, the constraints on the mixing parameters 
are
\begin{eqnarray}
 \label{bounds2} 
 0.203 \left| \Lambda_{\rm L}^{ij} + \Xi_{\rm L}^{ij} \Theta \right|^2 +
 0.176 \left| \Lambda_{\rm R}^{ij} + \Xi_{\rm R}^{ij} \Theta \right|^2
 &<& c_{l_i} {\rm \widetilde{B}}_{l_i l_j l_j \bar{l}_j},
\end{eqnarray}
where
$c_{l_i} = \left(
4 {\rm B} (l_i \rightarrow l_j \bar{\nu}_{l_j} \nu_{l_i}) \right)^{-1}$
and
\begin{equation}
 {\rm B} (l_i \rightarrow l_j \bar{\nu}_{l_j} \nu_{l_i}) = 
 \left\{
  \begin{array}{lcl}
   \rm B_{\rm \mu e \bar{\nu}_e \nu_\mu}     &\approx& 1.00             \\[2mm]
   \rm B_{\tau \mu \bar{\nu}_\mu \nu_\tau}   &=&       0.1735 \pm 0.0014\\[2mm]
   \rm B_{\rm \tau e \bar{\nu}_e \nu_\tau}   &=&       0.1783 \pm 0.0008 .
  \end{array}
 \right.
\end{equation}
 As in sec.~\ref{Zll}, the contribution of $\Theta$ in eq.~(\ref{bounds2}) is 
important when
\begin{equation}
 \Xi^{ij}_{\rm L} \Theta \gtrsim 
       \sqrt{\frac{c_{l_i}{\rm \widetilde{B}}_{l_i l_i \bar{l}_j}}{0.203}}
 \hspace{10mm} {\rm and} \hspace{10mm}
 \Xi^{ij}_{\rm R} \Theta \gtrsim 
       \sqrt{\frac{c_{l_i}{\rm \widetilde{B}}_{l_i l_i \bar{l}_j}}{0.176}}.
\end{equation}
 The bounds for $\Lambda^{\rm \tau e}_a + \Xi^{\rm \tau e}_a \Theta$ and 
$\Lambda^{\tau \mu}_a + \Xi^{\tau \mu}_a \Theta$ obtained from 
eq.~(\ref{bounds2}) are similar to those obtained from eq.~(\ref{bounds1}).
 For the $\rm \mu e$ case eq.~(\ref{bounds2}) is more stringent than 
eq.~(\ref{bounds1}). 
 The results of this section are resumed in table \ref{table2}.

\subsubsection{Constraints from $l_i \rightarrow l_j l_k \bar{l}_k$}
Assuming $m_{l_i} \gg m_{l_j}$, $m_{l_k}$, ignoring possible 
contributions from scalars and neglecting the tree level diagrams which 
involve simultaneously two FCNC vertices, the branching ratio 
${\rm B}(l_i \rightarrow l_j l_k \bar{l}_k)$ is
\begin{mathletters}
\begin{eqnarray}
 \frac{{\rm B} (l_i \rightarrow l_j l_k \bar{l}_k)}
      {{\rm B} (l_i \rightarrow l_j \bar{\nu}_{l_j} \nu_{l_i})} 
 &=& 
   \left(
    \left|
     g^{kk}_{\rm V}
    \right|^2 + 
    \left|
     g^{kk}_{\rm A}
    \right|^2
   \right) 
   \left(
    \left|
     g^{ij}_{\rm V}
    \right|^2 + 
    \left|
     g^{ij}_{\rm A}
    \right|^2
   \right)                        \nonumber \\[2mm]  
 && + \frac{M^2_{\rm Z}}{M^2_{\rm Z'}}
  2\Re e
  \left(
   g^{kk}_{\rm V} {g^{\prime kk}_{\rm V}}^\ast + 
   g^{kk}_{\rm A} {g^{\prime kk}_{\rm A}}^\ast
  \right)
  \left(
   g^{ij}_{\rm V} {g^{\prime ij}_{\rm V}}^\ast + 
   g^{ij}_{\rm A} {g^{\prime ij}_{\rm A}}^\ast
  \right)                            \nonumber \\[2mm] 
  && + \frac{M^4_{\rm Z}}{M^4_{\rm Z'}}
   \left(
    \left|
     g^{\prime kk}_{\rm V}
    \right|^2 + 
    \left|
     g^{\prime kk}_{\rm A}
    \right|^2
   \right) 
   \left(
    \left|
     g^{\prime ij}_{\rm V}
    \right|^2 + 
    \left|
     g^{\prime ij}_{\rm A}
    \right|^2
   \right) 
\label{explain1} \\[3mm]
&\simeq&
 4
  \left(
   \left|
    -\case{1}{2} + {\rm s_{\theta_w}^2}
   \right|^2 +
   \left|
    {\rm s_{\theta_w}^2}
   \right|^2 
  \right)
  \left(
   \left|
    \Lambda^{ij}_{\rm L} + \Xi^{ij}_{\rm L} \Theta 
   \right|^2 + 
   \left|
    \Lambda^{ij}_{\rm R} + \Xi^{ij}_{\rm R} \Theta 
   \right|^2
  \right)   
 + {\rm O}\left( \Theta^2\right),
\end{eqnarray} 
 where we made the same assumptions as in section~\ref{liljlj}.
 Using the experimental limits~\cite{prl73:1890}
\end{mathletters}
\begin{equation}
 {\rm B}_{l_i l_j l_k \bar{l}_k} \equiv
 {\rm B}(l_i \rightarrow l_j l_k \bar{l}_k ) = 
 \left\{
  \begin{array}{lcclcc}
   \rm B_{\tau e \mu \bar{\mu}}       &<& 3.6 \times 10^{-6}  &\equiv& 
                              \rm \widetilde{B}_{\tau e \mu \bar{\mu}}  \\[2mm]
   \rm B_{\tau \mu e \bar{e}}         &<& 3.4 \times 10^{-6}  &\equiv&
                              \rm \widetilde{B}_{\tau \mu e \bar{e}},
  \end{array}
 \right.
\end{equation}
the constraints on the mixing parameters are
\begin{eqnarray}
 \label{boundsijk} 
 0.126 
 \left(
  \left| 
   \Lambda_{\rm L}^{ij} + \Xi_{\rm L}^{ij} \Theta 
  \right|^2 +
  \left| 
   \Lambda_{\rm R}^{ij} + \Xi_{\rm R}^{ij} \Theta 
  \right|^2
 \right)
 &<& c_{l_i} {\rm \widetilde{B}}_{l_i l_j l_k \bar{l}_k}
\end{eqnarray}
which are not of interest in our analysis since they are somewhat 
weaker than those of eq.~(\ref{bounds2}).

\section{Some $\bbox{\rm SU(2)_L}$ representation for additional fermions}
\label{irreps}
 Some improvement on the above derived bounds for the mixing parameters 
may be obtained with information about the $\rm SU(2)_L$ transformation 
properties of the additional fermions and for this reason 
we analyze here a few simple $\rm SU(2)_L$ representations 
in which new additional charged leptons may appear.
In this analysis we will not consider any particular case for
the $\Xi^{ij}_a$ parameters, but we will assume that they are of O(1).
 What follows is valid for one or more additional families, independently 
of whether the extra families are fundamental or excited leptons in the 
context of composite models.

\subsection{No additional fermions}
 Equations~(\ref{bounds1}) and (\ref{bounds2}) are valid even if no 
additional charged leptons are present in the extended theory. 
 In this case 
${\sf F} = \bbox{\Lambda}_{\rm L} = \bbox{\Lambda}_{\rm R} = {\sf 0}$ 
and 
\begin{eqnarray}
 \bbox{\Xi}_a &=& 
 \left(
  \sf A^\dagger H_{\rm O} A 
 \right)_a .
\end{eqnarray}
 There are three subcases:
\begin{enumerate}
 \item ${\sf H}_{\rm O}$ is of the UFD type. Then Z does not couple to FCNC
       since $\left({\sf A^\dagger A}\right)_a = {\sf 1}$
       and therefore
 \begin{equation}
  \Xi_{\rm L}^{ij} = \Xi_{\rm R}^{ij} = 0 
  \hspace{4mm} {\rm for} \; i\neq j,
 \end{equation}
 \item ${\sf H}_{\rm O}$ is of the NUFD type. 
       Then there are two possibilities:
 \begin{enumerate}
  \item ${\sf A} = {\sf 1}$ (no mixing among the ordinary leptons).
        Then Z does not couple to FCNC.
  \item ${\sf A} \neq {\sf 1}$. Then Z couples to FCNC.
 \end{enumerate}
 \item ${\sf H}_{\rm O}$ is of the FCNC type. Then Z couples to FCNC.
\end{enumerate}
 Therefore when FCNC exists, eqs.~(\ref{bounds1}) and (\ref{bounds2}) read:
\begin{eqnarray}
\label{bounds3}
 \left| \Xi_{\rm L}^{ij} \Theta \right|^2 +
 \left| \Xi_{\rm R}^{ij} \Theta \right|^2
 &<& c {\rm \widetilde{B}}_{l_i \bar{l}_j} =
 \left\{
  \begin{array}{ccc}
   c {\rm \widetilde{B}}_{e \bar{\mu}}    &=& 3.2 \times 10^{-6} \\[2mm]
   c {\rm \widetilde{B}}_{e \bar{\tau}}   &=& 1.4 \times 10^{-5} \\[2mm]
   c {\rm \widetilde{B}}_{\mu \bar{\tau}} &=& 1.9 \times 10^{-5} 
  \end{array}
 \right.
\end{eqnarray}
and
\begin{eqnarray}
 \label{bounds4} 
 0.203 \left| \Xi_{\rm L}^{ij} \Theta \right|^2 +
 0.176 \left| \Xi_{\rm R}^{ij} \Theta \right|^2
 &=& c_{l_i} {\rm B}_{l_i l_j l_j \bar{l}_j} <
 \left\{
  \begin{array}{lclcl}
   c_\mu  {\rm \widetilde{B}}_{\mu e e \bar{e}}        &=& 0.25 \times 10^{-12} \\[2mm]
   c_\tau {\rm \widetilde{B}}_{\tau e e \bar{e}}       &=& 4.8  \times 10^{-6}  \\[2mm]
   c_\tau {\rm \widetilde{B}}_{\tau \mu \mu \bar{\mu}} &=& 2.7  \times 10^{-6} 
  \end{array} 
 \right.
\end{eqnarray}
respectively. All the constraints are for the product 
$\Xi_a^{ij} \Theta$ of the couplings of the light fermions to the Z$'$ 
and the Z-Z$'$ mixing angle. 
In particular $\Xi_a^{\rm e \mu}\Theta < 10^{-6}$.

\subsection{Sequential fermions}
\begin{mathletters}
\begin{eqnarray}
 \left.
  \begin{array}{ccc}
   t_{\rm 3EL} &=& -\frac{1}{2} \\[3mm]
   t_{\rm 3ER} &=& 0 
  \end{array}
 \right\}
 \hspace{4mm} &\Longrightarrow& \hspace{4mm}
 \bbox{\Lambda}_{\rm L} = \bbox{\Lambda}_{\rm R} = {\sf 0}, \\[3mm]
 \bbox{\Xi}_a &=& 
 \left(
  \sf A^\dagger H_{\rm O} A  + F^\dagger H_{\rm E} F 
 \right)_a .
\end{eqnarray}
 The situation is the same as that of no additional fermions.
Eqs.~(\ref{bounds3}) and (\ref{bounds4}) hold with the only
difference that when $\sf F \neq 0$ then $\bbox{\Xi}_a$ contains 
the $\left({\sf F^\dagger H_{\rm E} F }\right)_a$ contribution.
 As in the previous case the strongest constraint is for  
$\Xi_a^{\rm e \mu}\Theta < 10^{-6}$.
\end{mathletters}

\subsection{Vector singlets}
\begin{mathletters}
\begin{eqnarray}
 \left.
  \begin{array}{ccc}
   t_{\rm 3EL} &=& 0 \\[3mm]
   t_{\rm 3ER} &=& 0 
  \end{array}
 \right\}
 \hspace{4mm} &\Longrightarrow& \hspace{4mm}
 \bbox{\Lambda}_{\rm R} = {\sf 0} \hspace{4mm} 
 \bbox{\Lambda}_{\rm L} = \case{1}{2}
 \left(
  {\sf F^\dagger F}
 \right)_{\rm L}, \\[3mm]
 \bbox{\Xi}_a &=& 
 \left(
  \sf A^\dagger H_{\rm O} A  + F^\dagger H_{\rm E} F 
 \right)_a .
\end{eqnarray}
Therefore eqs.~(\ref{bounds1}) and (\ref{bounds2}) now read
\end{mathletters}
\begin{eqnarray}
\label{bounds5}
 \left| \Lambda_{\rm L}^{ij} + \Xi_{\rm L}^{ij} \Theta \right|^2 +
 \left|                            \Xi_{\rm R}^{ij} \Theta \right|^2
 &<& c {\rm \widetilde{B}}_{l_i \bar{l}_j} 
\end{eqnarray}
and
\begin{eqnarray}
 \label{bounds6} 
 0.203 \left| 
         \Lambda_{\rm L}^{ij} + \Xi_{\rm L}^{ij} \Theta
        \right|^2 +
 0.176 \left| 
         \Xi_{\rm R}^{ij} \Theta
        \right|^2
 &<& c_{l_i} {\rm \widetilde{B}}_{l_i l_j l_j \bar{l}_j} 
\end{eqnarray}
respectively.
 The contribution to FCNC from the ordinary--exotic fermion mixing is
only left handed.
If $\Xi_a^{ij} \sim {\rm O(1)}$, then the stringent bounds on 
$\Theta$, consequence of $\Xi_{\rm R}^{\rm e\mu} \Theta < 10^{-6}$,
imply an equally stringent bound on $\Lambda^{\rm e \mu}_{\rm L}$. 

\subsection{Vector doublets (homodoublets)}
\begin{mathletters}
\begin{eqnarray}
 \left.
  \begin{array}{ccc}
   t_{\rm 3EL} &=& -\frac{1}{2} \\[3mm]
   t_{\rm 3ER} &=& -\frac{1}{2} 
  \end{array}
 \right\}
 \hspace{4mm} &\Longrightarrow& \hspace{4mm}
 \bbox{\Lambda}_{\rm L} = {\sf 0}
 \hspace{4mm} 
 \bbox{\Lambda}_{\rm R} = -\case{1}{2} 
 \left(
  \sf F^\dagger F
 \right)_{\rm R} ,\\[3mm]
 \bbox{\Xi}_a &=& 
 \left(
  \sf A^\dagger H_{\rm O} A  + F^\dagger H_{\rm E} F 
 \right)_a .
\end{eqnarray}
 Thus eqs.~(\ref{bounds1}) and (\ref{bounds2}) now read
\end{mathletters}
\begin{eqnarray}
\label{bounds7}
 \left|                         
  \Xi_{\rm L}^{ij} \Theta 
 \right|^2 +
 \left| 
  \Lambda_{\rm R}^{ij} + \Xi_{\rm R}^{ij} \Theta 
 \right|^2
 &<& c {\rm \widetilde{B}}_{l_i \bar{l}_j} 
\end{eqnarray}
and
\begin{eqnarray}
 \label{bounds8} 
 0.203 \left| 
         \Xi_{\rm L}^{ij} \Theta 
        \right|^2 +
 0.176 \left|
         \Lambda_{\rm R}^{ij} + \Xi_{\rm R}^{ij} \Theta 
        \right|^2
 &<& c_{l_i} {\rm \widetilde{B}}_{l_i l_j l_j \bar{l}_j}.
\end{eqnarray}
The contribution to FCNC from the ordinary--exotic fermions mixing is
only right handed. The conclusions are the same as in the previous
case with $\rm L \leftrightarrow R$.

\subsection{Mirror fermions}
\begin{mathletters}
\begin{eqnarray}
 \left.
  \begin{array}{ccc}
   t_{\rm 3EL} &=& 0 \\[3mm]
   t_{\rm 3ER} &=& -\frac{1}{2} 
  \end{array}
 \right\}
 \hspace{4mm} &\Longrightarrow& \hspace{4mm}
 \bbox{\Lambda}_{\rm L} = \case{1}{2}
 \left(
  \sf F^\dagger F
 \right)_{\rm L} 
 \hspace{4mm} 
 \bbox{\Lambda}_{\rm R} = -\case{1}{2}
 \left(
  \sf F^\dagger F
 \right)_{\rm R}, \\[3mm]
 \bbox{\Xi}_a &=& 
 \left(
  \sf A^\dagger H_{\rm O} A  + F^\dagger H_{\rm E} F 
 \right)_a .
\end{eqnarray}
 Hence eqs.~(\ref{bounds1}) and (\ref{bounds2}) are unchanged
\end{mathletters}
\begin{eqnarray}
\label{bounds9}
 \left| 
  \Lambda_{\rm L}^{ij} + \Xi_{\rm L}^{ij} \Theta 
 \right|^2 +
 \left|
  \Lambda_{\rm R}^{ij} + \Xi_{\rm R}^{ij} \Theta 
 \right|^2
 &<& c {\rm \widetilde{B}}_{l_i \bar{l}_j} 
\end{eqnarray}
and
\begin{eqnarray}
 \label{bounds10} 
 0.203 \left|
         \Lambda_{\rm L}^{ij} + \Xi_{\rm L}^{ij} \Theta 
        \right|^2 +
 0.176 \left|
         \Lambda_{\rm R}^{ij} + \Xi_{\rm L}^{ij} \Theta 
        \right|^2
 &<& c_{l_i} {\rm \widetilde{B}}_{l_i l_j l_j \bar{l}_j} .
\end{eqnarray}
The contribution to FCNC from the ordinary--exotic fermions mixing is
both left and right handed.
As a consequence there are no stringent bounds on $\Theta$ and the limits
on $\Lambda_a^{ij}$ and $\Theta$ are strongly correlated.

\subsection{Self conjugated triplets}
\begin{mathletters}
\begin{eqnarray}
 \left.
  \begin{array}{ccc}
   t_{\rm 3EL} &=& -1 \\[3mm]
   t_{\rm 3ER} &=& -1 
  \end{array}
 \right\}
 \hspace{4mm} &\Longrightarrow& \hspace{4mm}
 \bbox{\Lambda}_{\rm L} = -\case{1}{2}
 \left(
  \sf F^\dagger F
 \right)_{\rm L} 
 \hspace{4mm} 
 \bbox{\Lambda}_{\rm R} = -
 \left(
  \sf F^\dagger F
 \right)_{\rm R} ,\\[3mm]
 \bbox{\Xi}_a &=& 
 \left(
  \sf A^\dagger H_{\rm O} A  + F^\dagger H_{\rm E} F 
 \right)_a .
\end{eqnarray}
 Hence eqs.~(\ref{bounds1}) and (\ref{bounds2}) are unchanged
\end{mathletters}
\begin{eqnarray}
\label{bounds11}
 \left| 
  \Lambda_{\rm L}^{ij} + \Xi_{\rm L}^{ij} \Theta 
 \right|^2 +
 \left| 
  \Lambda_{\rm R}^{ij} + \Xi_{\rm R}^{ij} \Theta 
 \right|^2
 &<& c {\rm \widetilde{B}}_{l_i \bar{l}_j} 
\end{eqnarray}
and
\begin{eqnarray}
 \label{bounds12} 
 0.203 \left| 
         \Lambda_{\rm L}^{ij} + \Xi_{\rm L}^{ij} \Theta 
        \right|^2 +
 0.176 \left| 
         \Lambda_{\rm R}^{ij} + \Xi_{\rm R}^{ij} \Theta 
        \right|^2
 &<& c_{l_i} {\rm \widetilde{B}}_{l_i l_j l_j \bar{l}_j} .
\end{eqnarray}
As in the previous case the contribution to FCNC from the ordinary--exotic 
fermions mixing is both left and right handed.
As a consequence there are no stringent bounds on $\Theta$ and the limits
on $\Lambda_a^{ij}$ and $\Theta$ are strongly correlated.

\section{Conclusions}
 In a model independent way we obtained bounds for the strength of the 
FCNC, $\left(\bbox{\Lambda} + \bbox{\Xi} \Theta \right)_a$, 
in the ordinary charged--leptons sector, produced both by the 
ordinary--exotic fermion mixing, $\Lambda^{ij}_a$, and by the 
Z--Z$'$ mixing, $\Theta$. 
 Giving that the experimental bounds on the decay 
$\rm \mu \rightarrow ee \bar{e}$ are more stringent than those for the 
FC decays of the $\tau$ into three charged leptons and of the Z into 
two charged leptons, the bounds on the $\mu$--e coupling of the Z are 
stronger than those on the $\tau$--e and $\tau$--$\mu$ couplings.
 We have shown also that in some cases, when the $\rm SU(2)_L$ 
representation of the additional fermions is relatively simple, the 
bounds may be refined.
 In other cases there may be a strong correlation between 
$\Theta$ and $\Lambda_a^{ij}$ and then it is
not safe to take the limit $\Theta \rightarrow 0$. 
 In the same way, if one consider specific extended models, 
e.g.~\cite{prd32:306,npb437:491,plb187:303,plb188:91,plb197:418,%
plb199:432,npb477:321,plb374:80,prd51:6474},
some additional statements may be drawn on the $\Xi^{ij}_a$.
 In this work we have concentrated our attention to LFV in decay processes.
 On the other hand, there may be LFV processes of a different 
type~\cite{desy96:260} which will certainly put additional constraints on
the LFV parameters.

\section*{Acknowledgements}
This work was partially supported by CONACyT in Mexico. 
One of us (A.Z.) acknowledges the hospitality of Prof. J. Bernabeu and of 
the Theory Group at the University of Valencia as well as the financial 
support, during the 1995-1996 sabbatical 
leave, of Direcci\'on General de Investigaci\'on Cient\'\i fica y 
T\'ecnica (DGICYT) of the Ministry of Education and Science of Spain.



\begin{table}[ht]
\begin{center}
\begin{tabular}{c}
 \hline 
 \hline \\
  Limits from ${\rm Z} \rightarrow l_i \bar{l}_j$ \\ 
$
 \left| \Lambda_{\rm L}^{ij} + \Xi_{\rm L}^{ij} \Theta \right|^2 +
 \left| \Lambda_{\rm R}^{ij} + \Xi_{\rm R}^{ij} \Theta \right|^2
 \;<\; 1.87 \, {\rm \widetilde{\rm B}}_{l_i \bar{l}_j} \;=\;
 1.87 \,\times\,
 \left\{
  \begin{array}{ccl}
   1.7 \times 10^{-6} &\equiv& \rm \widetilde{B}_{e \bar{\mu}}  \\
   7.3 \times 10^{-6} &\equiv& \rm \widetilde{B}_{e \bar{\tau}} \\
   1.0 \times 10^{-5} &\equiv& \rm \widetilde{B}_{\mu \bar{\tau}},
  \end{array}
 \right. 
$ 
 \\ [4ex]
 \hline \\
 Mixing e--$\mu$  \\[1ex]
$
 \left| \Lambda_{\rm L}^{\rm e\mu} + \Xi_{\rm L}^{\rm e\mu} \Theta \right|^2 +
 \left| \Lambda_{\rm R}^{\rm e\mu} + \Xi_{\rm R}^{\rm e\mu} \Theta \right|^2
 \;<\; 3.2 \times 10^{-6} 
 \hspace{3ex} \Longrightarrow \hspace{3ex} 
 \left| \Lambda_a^{\rm e\mu} + \Xi_a^{\rm e\mu} \Theta \right|
 \, < \, 1.8 \times 10^{-3}
$
 \\ [2ex]
 \hline \\
 Mixing e--$\tau$  \\[1ex]
$
 \left| \Lambda_{\rm L}^{\rm e\tau} + \Xi_{\rm L}^{\rm e\tau} \Theta \right|^2 +
 \left| \Lambda_{\rm R}^{\rm e\tau} + \Xi_{\rm R}^{\rm e\tau} \Theta \right|^2
 \;<\; 1.4 \times 10^{-5} 
 \hspace{3ex} \Longrightarrow \hspace{3ex} 
 \left| \Lambda_a^{\rm e\tau} + \Xi_a^{\rm e\tau} \Theta \right|
 \, < \, 3.7 \times 10^{-3}
$
 \\[2ex]
 \hline \\
 Mixing $\mu$--$\tau$  \\[1ex]
$
 \left| \Lambda_{\rm L}^{\mu\tau} + \Xi_{\rm L}^{\mu\tau} \Theta \right|^2 +
 \left| \Lambda_{\rm R}^{\mu\tau} + \Xi_{\rm R}^{\mu\tau} \Theta \right|^2
 \;<\; 1.9 \times 10^{-5} 
 \hspace{3ex} \Longrightarrow \hspace{3ex} 
 \left| \Lambda_a^{\mu\tau} + \Xi_a^{\mu\tau} \Theta \right|
 \, < \, 4.3 \times 10^{-3}
$
 \\[2ex]
 \hline
 \hline
\end{tabular}
\end{center}
\caption{Bounds from the process ${\rm Z} \rightarrow l_i \bar{l}_j$}
\label{table1}
\end{table}

\begin{table}[ht]
\begin{center}
\begin{tabular}{c}
 \hline 
 \hline\\
  Limits from $l_i \rightarrow l_j l_j \bar{l}_j$ \\[2ex]
$
 0.203 \left| \Lambda_{\rm L}^{ij} + \Xi_{\rm L}^{ij} \Theta \right|^2 +
 0.176 \left| \Lambda_{\rm R}^{ij} + \Xi_{\rm R}^{ij} \Theta \right|^2
 \;<\; c_{l_i} \, {\rm \widetilde{\rm B}}_{l_i l_j l_j \bar{l}_j} \;=\;
 c_{l_i} \,\times\,
 \left\{
  \begin{array}{ccl}
   1.0 \times 10^{-12} &\equiv& \rm \widetilde{B}_{\mu  ee\bar{e}}   \\
   3.3 \times 10^{-6}  &\equiv& \rm \widetilde{B}_{\tau ee\bar{e}}   \\
   1.9 \times 10^{-6}  &\equiv& \rm \widetilde{B}_{\tau \mu\mu\bar{\mu}}, 
  \end{array}
 \right.
$ 
 \\[4ex]
 \hline \\
 Mixing e--$\mu$  \\[1ex]
$
 0.203 \left| \Lambda_{\rm L}^{\rm e\mu} + \Xi_{\rm L}^{\rm e\mu} \Theta \right|^2 +
 0.176 \left| \Lambda_{\rm R}^{\rm e\mu} + \Xi_{\rm R}^{\rm e\mu} \Theta \right|^2
 \;<\; 0.25 \times 10^{-12} 
$
 \\[2ex]
$
 \left| \Lambda_{\rm L}^{\rm e\mu} + \Xi_{\rm L}^{\rm e\mu} \Theta \right|
 \, < \, 1.1 \times 10^{-6} \hspace{4ex} 
 \left| \Lambda_{\rm R}^{\rm e\mu} + \Xi_{\rm R}^{\rm e\mu} \Theta \right|
 \, < \, 1.2 \times 10^{-6} 
$
 \\
 \\[2ex]
 \hline \\
 Mixing e--$\tau$  \\[1ex]
$
 0.203 \left| \Lambda_{\rm L}^{\rm e\tau} + \Xi_{\rm L}^{\rm e\tau} \Theta \right|^2 +
 0.176 \left| \Lambda_{\rm R}^{\rm e\tau} + \Xi_{\rm R}^{\rm e\tau} \Theta \right|^2
 \;<\; 4.8 \times 10^{-6} 
$
 \\[2ex]
$
  \left| \Lambda_{\rm L}^{\rm e\tau} + \Xi_{\rm L}^{\rm e\tau} \Theta \right|
  \, < \, 4.9 \times 10^{-3} \hspace{4ex}
  \left| \Lambda_{\rm R}^{\rm e\tau} + \Xi_{\rm R}^{\rm e\tau} \Theta \right|
  \, < \, 5.2 \times 10^{-3} 
$
 \\
 \\[2ex]
 \hline \\
 Mixing $\mu$--$\tau$  \\[1ex]
$
 0.203 \left| \Lambda_{\rm L}^{\mu\tau} + \Xi_{\rm L}^{\mu\tau} \Theta \right|^2 +
 0.176 \left| \Lambda_{\rm R}^{\mu\tau} + \Xi_{\rm R}^{\mu\tau} \Theta \right|^2
 \;<\; 2.7 \times 10^{-6} 
$
 \\[2ex]
$
  \left| \Lambda_{\rm L}^{\mu\tau} + \Xi_{\rm L}^{\mu\tau} \Theta \right|
  \, < \, 3.6 \times 10^{-3} \hspace{4ex}
  \left| \Lambda_{\rm R}^{\mu\tau} + \Xi_{\rm R}^{\mu\tau} \Theta \right|
  \, < \, 3.9 \times 10^{-3} 
$
 \\
 \\[2ex]
\hline
\hline
\end{tabular}
\end{center}
\caption{Bounds from the process $l_i \rightarrow l_j l_j\bar{l}_j$}
\label{table2}
\end{table}

\end{document}